# Interfacial Free Energy as the Key to the Pressure-Induced Deceleration of Ice Nucleation

Jorge R. Espinosa,[1] Alberto Zaragoza,[1] Pablo Rosales-Pelaez,[1] Caridad Navarro,[1] Chantal Valeriani,[1,2]
Carlos Vega,[1] and Eduardo Sanz[1]
[1]*Departamento de Quimica Fisica I, Facultad de Ciencias Quimicas, Universidad Complutense de Madrid, 28040 Madrid, Spain*
[2]*Departamento de Fisica Aplicada I, Facultad de Ciencias Fisicas, Universidad Complutense de Madrid, 28040 Madrid, Spain*

The avoidance of water freezing is the holy grail in the cryopreservation of biological samples, food, and organs. Fast cooling rates are used to beat ice nucleation and avoid cell damage. This strategy can be enhanced by applying high pressures to decrease the nucleation rate, but the physics behind this procedure has not been fully understood yet. We perform computer experiments to investigate ice nucleation at high pressures consisting in embedding ice seeds in supercooled water. We find that the slowing down of the nucleation rate is mainly due to an increase of the ice $I$-water interfacial free energy with pressure. Our work also clarifies the molecular mechanism of ice nucleation for a wide pressure range. This study is not only relevant to cryopreservation, but also to water amorphization and climate change modeling.

In 1975 Kanno *et al.* [1] studied experimentally homogeneous ice nucleation (i.e., in the absence of surfaces and impurities) for pressures up to 2000 bar. By measuring the temperature at which microscopic emulsified water drops freeze when cooled at a rate of a few Kelvin per minute they established the so-called homogeneous nucleation line (HNL), whose slope is negative and larger than that of the melting line. Thus, they found that whereas water remains liquid for temperatures down to -38 Celsius at ambient pressure, at high pressures it is possible to have liquid water at temperatures as low as -92 Celsius. Therefore, applying pressure significantly increases the range of temperatures at which liquid water may exist. This important experimental result is the basis of state-of-the-art coolers used for the preservation of biological samples [2,3]. Despite its importance, the experiment by Kanno *et al.* has long remained unexplained.

Using a combination of simulation methods to obtain nucleation rates and interfacial free energies [4–6] we find that the decelerating effect of pressure on ice nucleation is due to the increase of the ice $I$-water interfacial free energy. This conclusion is supported by both studied models: TIP4P/Ice [7] (main text) and mW [8] [Supplemental Material (SM) [9] ]. These findings have implications in cryopreservation [2], water vitrification [10,11], and planetary and atmospheric science [12,13]. The importance and broad impact of simulation studies of crystal nucleation, notably for water, are highlighted in a very recent review [14].

To perform this research we use the seeding method [4,5,15,16], which is a computer experiment that consists in embedding an ice cluster in the supercooled fluid and following its evolution in a molecular dynamics simulation at constant pressure and temperature (see SM [9] for simulation details), to determine the critical cluster size. In Fig. 1(a) we show the number of molecules in the critical cluster, $N_c$, as a function of the supercooling, $\Delta T$ (the difference between the melting temperature, $T_m$, and the temperature of interest), for 1 and 2000 bars. We insert spherical ice Ih clusters (according to our previous work we do not expect differences if ice Ic had been considered instead [20]). Clearly, the supercooling at which a certain cluster size is critical increases with pressure. This result already points to the experimental observation that pressure hinders ice nucleation.

In order to know for certain whether nucleation is slowed down with pressure one needs to evaluate the nucleation rate, $J$ (the number of critical clusters that nucleate per unit time and volume). Classical nucleation theory (CNT) [21–23] gives the following expression for the nucleation rate:

$$J = A\rho_f \exp[-\Delta G_c/(k_B T)], \qquad (1)$$

where $A$ is a kinetic prefactor, $\rho_f$ the fluid number density, and $\Delta G_c/(k_B T)$ the free energy penalty associated to the appearance of the critical cluster [$\rho_f \exp[-\Delta G_c/(k_B T)]$ is then the number density of critical clusters]. In turn, $\Delta G_c/(k_B T)$ is given by two competing terms, a favorable one due to the lowering of the chemical potential when a liquid molecule incorporates into the crystal, and an unfavorable one due to the cost of creating an interface between the liquid and the crystal,

$$\Delta G(N) = -N|\Delta\mu| + S(N)\gamma, \qquad (2)$$

where $N$ is the number of molecules in the cluster, $|\Delta\mu|$ the ice-water chemical potential difference, $S$ the area of the cluster surface, and $\gamma$ the water-ice interfacial free energy. $\Delta G_c$ is derived by maximizing Eq. (2) with respect to $N$,

$$\Delta G_c = (N_c|\Delta\mu|)/2, \qquad (3)$$

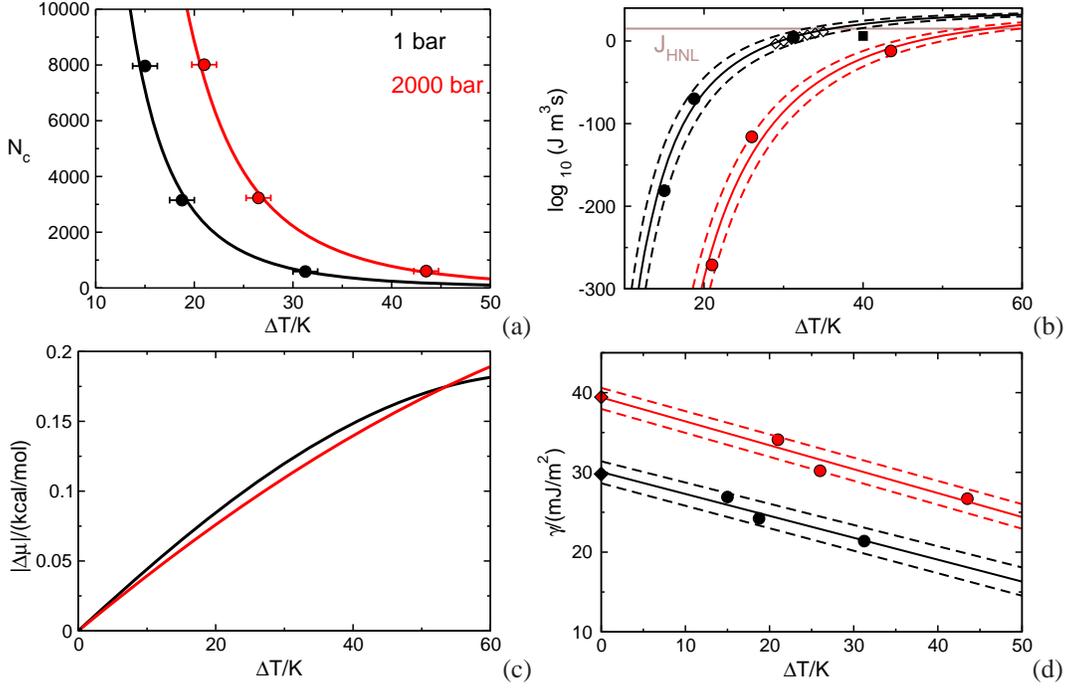

FIG. 1. Plotted for 1 (black) and 2000 bar (red) as a function of the supercooling: (a) number of particles in the ice Ih critical cluster; (b) decimal logarithm of the ice nucleation rate; (c) absolute value of the chemical potential difference between ice Ih and the liquid from thermodynamic integration [17]; (d) water-ice Ih interfacial free energy. In (a), (b), and (d) solid circles correspond to data obtained from inserted clusters and solid lines to fits [CNT-like in (a) and (b) as explained in Ref. [17], and linear in (d)]. Diamonds in (d) correspond to the direct calculation of $\gamma$ at coexistence (flat interface) from the MI method [6] (error is the size of the symbol). Dashed lines in (b) and (d) correspond to the boundaries of our statistical error (see SM [9]). The black solid square in (b) corresponds to the forward flux sampling calculation performed at 1 bar for TIP4P/Ice in Ref. [18]. Empty diamonds in (b) correspond to experimental measurements of $J$ at 1 bar [19].

where $N_c$ is the number of molecules in the critical cluster. The kinetic prefactor is given by

$$A = \sqrt{\frac{|\Delta\mu|}{6\pi k_B T N_c}} f^+, \quad (4)$$

where $f^+$ is the attachment rate of molecules to the critical cluster, calculated as the diffusion coefficient of the number of particles in the critical cluster [24]: $f^+ = \langle [N(t) - N_c]^2/(2t) \rangle$.

In summary, to compute $J$ one needs to obtain $N_c$, $|\Delta\mu|$, $\rho_f$, and $f^+$ by simulation and combine them as in the equations above. Details on the way these variables are obtained are given in our previous work [4,5,17]. The obtained values of $J$ as a function of the supercooling are plotted as filled circles in Fig. 1(b). The results for 1 bar have an improved accuracy with respect to our previous work [4,17]. The thermodynamic driving force for nucleation, $|\Delta\mu(T)|$, is plotted in Fig. 1(c).

To fit our $J$ data we first obtain $\gamma$ for the inserted clusters using the following CNT expression:

$$\gamma = \left(\frac{3N_c \rho_s^2 |\Delta\mu|^3}{32\pi}\right)^{1/3}, \quad (5)$$

where $\rho_s$ is the number density of the cluster. The values thus obtained are shown with circles in Fig. 1(d). We also compute $\gamma$ at coexistence ($\Delta T = 0$) by means of the mold integration (MI) technique [6]. The results are shown with diamonds in Fig. 1(d) (the data for 1 bar were recently obtained by us in Ref. [25] and those for 2000 bar have been calculated in this work). By fitting $\gamma(T)$ to a straight line (combining both seeding and MI data), and $|\Delta\mu(T)|$ and $\rho_s(T)$ to a second degree polynomial, we obtain the dependency of $\Delta G_c$ with $T$ as

$$\frac{\Delta G_c(T)}{k_B T} = \frac{16\pi[\gamma(T)]^3}{3[\rho_s(T)]^2 |\Delta\mu(T)|^2 k_B T}. \quad (6)$$

On the other hand, we obtain $f^+(T)$ as described in Refs. [4,5,17] and combine it with $\Delta G_c(T)$ to obtain $J(T)$ via Eq. (1). The fits thus obtained are shown with solid curves in Fig. 1(b). The dashed lines correspond to the boundaries of our statistical error, which is mainly due to the determination of the temperature for which the inserted clusters become critical (see SM [9] for a detailed discussion on the error boundaries). We include experimental data, available only at 1 bar, in Fig. 1(b) (empty diamonds). For the sake of clarity we only show data from one group

[19] in the figure but, within the uncertainty of our calculations, TIP4P/Ice predictions agree with most published measurements of $J$ [26–31]. It is clear from Fig. 1(b) that the supercooling required to get a certain $J$ increases with pressure, in qualitative agreement with the experimental trend [1,32].

We now aim at predicting the location of the HNL with the employed water model. In the experimental work where the HNL was reported, microscopic drops of volume $V$ were cooled at a rate of $v = 3$ K/min until ice formation was observed [1]. The time, $\tau$, the sample spends at a given temperature interval, $\Delta T$, is $\Delta T/v$. Assuming that $J$ is roughly constant in every temperature interval (small $\Delta T$), the time needed for ice to nucleate when the droplet is at temperature $T \pm \Delta T/2$ is $\tau_N(T) = 1/[VJ(T)]$. While $\tau$ is constant, $\tau_N(T)$ decreases upon cooling. When $\tau_N(T)$ equals $\tau$ ice formation is observed. Thus, by equating both times and taking $\Delta T = 1$ K (the typical experimental temperature uncertainty) one can estimate the rate at which ice nucleates in Kanno et al.'s experiments as $J_{HNL} = v/V$. This gives $J_{HNL} \approx 10^{15}$ m$^{-3}$ s$^{-1}$ for drops with 5 $\mu$m radius. The homogeneous nucleation temperature at 1 bar is -38 °C [1]. At such thermodynamic conditions, the measured nucleation rate is $J = 10^{15}$ m$^{-3}$ s$^{-1}$ [28], which is consistent with the estimate above for $J_{HNL}$. Since neither $v$ nor $V$ is changed with pressure in the experiments by Kanno et al. [1], the HNL is an isonucleation rate line with $J = 10^{15}$ m$^{-3}$ s$^{-1}$.

Therefore, to obtain simulation predictions of the HNL we simply read from the fits given in Fig. 1(b) the temperature corresponding to $J_{HNL}$ [given by the cross between the horizontal line and the solid curves in Fig. 1(b)]. The points thus obtained are shown as red diamonds in Fig. 2(a). Their error bar is given by the cross between the horizontal line and the dashed curves in Fig. 1(b). The agreement between the experiment and the model is almost quantitative. According to the predictions by the TIP4P/Ice model the critical cluster contains between 300 (1 bar) and 200 (2000 bar) particles along the HNL. We show a snapshot of a critical cluster in the HNL at 1 bar in Fig. 2(b).

The question now is why nucleation is slowed down by pressure at constant supercooling. The natural logarithm of the nucleation rate is given by $\ln J = \ln(\rho_f A) - \Delta G_c/(k_B T)$. In Fig. 3(a) we plot the difference in $\ln J$ and in $-\Delta G_c/(k_B T)$ between both studied pressures as a function of $\Delta T$. Clearly, the decrease of $\ln J$ with pressure at constant $\Delta T$ can be essentially ascribed to an increase of $\Delta G_c/(k_B T)$. In order to have a quantitative assessment of which variable contributes most to such an increase we show in Fig. 3(b) the ratio between $\Delta G_c/(k_B T)$ at 2000 and 1 bar alongside all the factors that contribute to such a ratio [according to Eq. (6)]. $\Delta G_c/(k_B T)$ itself roughly increases by a factor of 2.5 (black curve). The $\rho_s$ factor (green curve) has a minor contribution due to the solid incompressibility [being smaller than 1 because $\rho_s$ increases with pressure and appears in the denominator of Eq. (6)]. The $T$ factor (red curve) increases $\Delta G_c/(k_B T)$ roughly by ten percent (slowing nucleation down) due to the decrease of the absolute temperature when pressure increases at constant $\Delta T$, as a consequence of the negative slope of the melting line. The $|\Delta\mu|$ factor (blue curve) raises $\Delta G_c/(k_B T)$ by 25 percent at low supercooling and slightly lowers it at deep supercooling. In fact, Fig. 1(c) shows that, for a given $\Delta T$, increasing the pressure first decreases $|\Delta\mu|$ and then increases it at high supercooling. The $\gamma$ factor (purple curve) multiplies by more than two $\Delta G_c/(k_B T)$ for all the supercooling range.

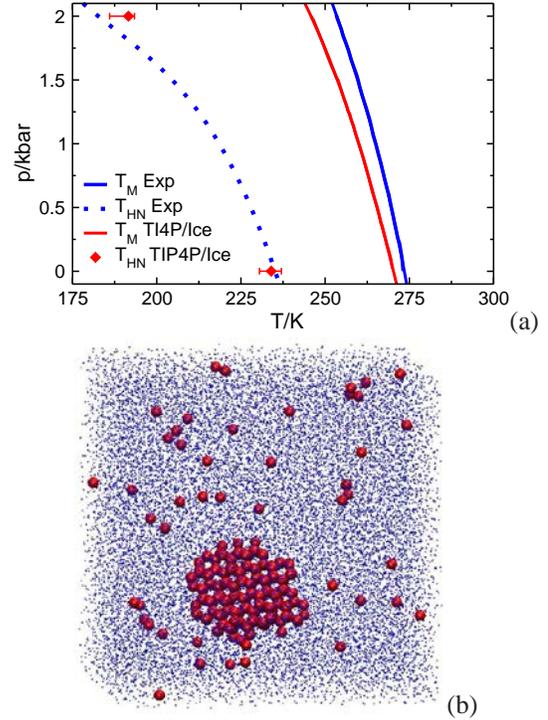

FIG. 2. (a) The experimental HNL (dotted blue) has a larger negative slope than the melting line (solid blue). The TIP4P/Ice model, in red, captures this effect almost quantitatively. (b) Snapshot of a typical critical cluster embedded in the supercooled fluid at the experimental HNL at 1 bar. Particles with an icelike environment are shown in red and those with a liquidlike environment in blue. The size of liquidlike particles has been scaled down to make the ice cluster visible.

Therefore, the variable that accounts to a greater extent for the effect of pressure on ice nucleation is the water-ice interfacial free energy, $\gamma$. In Fig. 1(d) we show how $\gamma$ notably increases with pressure for a given supercooling. Since it appears as a third power in the numerator of Eq. (6) its effect in raising the nucleation barrier (and therefore decreasing the nucleation rate) is quite important. Therefore, the increase of $\gamma$ with pressure is the key to understanding the slope of the HNL.

The dependence of $\gamma$ with pressure is totally unknown experimentally. In fact, there is not even a consensus for the experimental value of $\gamma$ at ambient pressure (there are reported values ranging from 25 to 35 mJ [33]). The increase of the solid-liquid interfacial free energy along the coexistence line cannot be seen as a water anomaly since it has also been reported for the Lennard-Jones system [34].

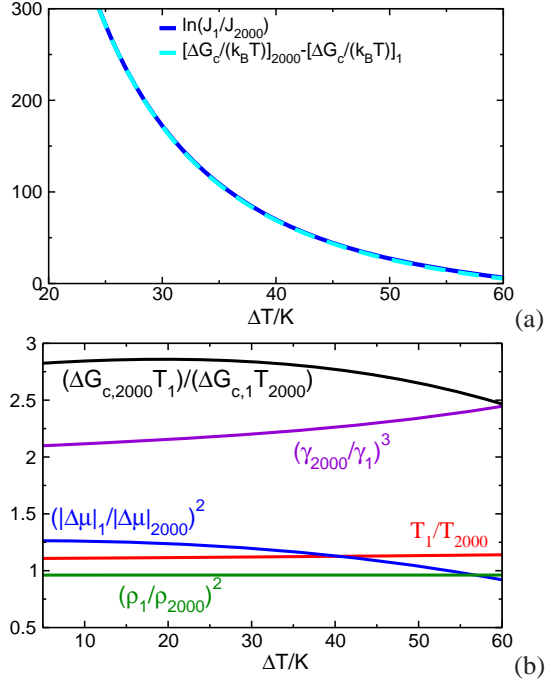

FIG. 3. (a) Difference in $\ln J$ and $-\Delta G_c/(k_B T)$ between 2000 and 1 bar as a function of the supercooling. (b) Black curve: factor by which $\Delta G_c/(k_B T)$ increases in raising the pressure from 1 to 2000 bar as a function of the supercooling. Colored curves: different factors that contribute to such an increase, as indicated in the figure. The product of the colored curves gives the black curve.

It is tempting to speculate that the observed increase of $\gamma$ with pressure is due to the pressure-induced breakage of hydrogen bonds in the liquid phase, which is an effect that can be also observed by adding salt [35,36]. In fact, the diffusion coefficient of water, contrary to that of most liquids, increases with pressure. By breaking its hydrogen bonds, the liquid decreases its structural resemblance to ice and the interfacial free energy between both phases increases. This reasoning could also justify the temperature dependence of $\gamma$ [37] (the liquid forms more hydrogen bonds on cooling and $\gamma$ decreases).

The slope of the linear fits to $\gamma(\Delta T)$ in Fig. 1(d) provides an estimate of interfacial entropy $s_{int} = -(d\gamma/dT) = [d\gamma/d(\Delta T)]$, which is $-0.27 \pm 0.05$ and $-0.30 \pm 0.05$ mJ/(m$^2$ K) for 1 and 2000 bar, respectively. The entropic contribution to $\gamma$, i.e., $-T_m s_{int}$, is, within our error bar, $73 \pm 10$ mJ/m$^2$ for both pressures. The interfacial enthalpy can be obtained as $h_{int} = \gamma + T_m s_{int}$, which gives $-43 \pm 11$ and $-34 \pm 11$ mJ/m$^2$ for 1 and 2000 bar, respectively. Two comments are due in view of these figures: (i) in accordance with previous works [4,38], we observe that creating an ice-water interface is enthalpically favorable and entropically unfavorable; (ii) our data suggest that $\gamma$ increases with pressure because forming the interface becomes enthalpically less favorable [this should be taken with caution, though, due to the large error bar in the slope of $\gamma(\Delta T)$].

In the mW water model the simulated HNL is parallel to the ice 0 melting line [39]. This has been interpreted as an evidence for ice 0 being the first crystal form to appear from the fluid [39]. Within this hypothesis, the mediation of ice 0 explains the decelerating effect of pressure on ice nucleation. This new scenario could dramatically change the interpretation of all nucleation rate measurements to date, based on the assumption that ice nucleates through an alternating stack of ice $I$ polymorphs [40–42]. The ice 0 hypothesis of Ref. [39] implicitly assumes that $\Delta\mu$, $T$, $\rho_s$, and $\gamma$ are independent of pressure for a given $\Delta T$. We show in this Letter that this is not the case, particularly so for $\gamma$. In the SM [9] we argue that ice 0 is not involved in ice nucleation with the same water model used in Ref. [39]. Thus, our work not only explains the effect of pressure on ice nucleation, but also provides strong evidence regarding molecular aspects of the ice nucleation mechanism such as the structure (ice $I$) and shape (spherical) of the critical cluster.

The seeding method here used is approximate and relies on the validity of CNT. Therefore, one has to perform consistency checks to test the validity of the calculations. One such test consists in extrapolating the seeding results for $\gamma$ to $\Delta T = 0$, where $\gamma$ can be independently and directly computed using a rigorous method such as MI. In Fig. 1(d) we show that the test is passed for TIP4P/Ice at both pressures. In the SM [9] we show that this point is also confirmed for the mW model. Another way to check the validity of seeding is to compare its results for the nucleation rate with those obtained by means of methods that do not rely on CNT [5]. Unfortunately, the nucleation rate predicted by us with seeding for TIP4P/Ice at 1 bar and $\Delta T = 40$ K is not consistent with a recent calculation by Haji-Akbari et al. using forward flux sampling (FFS) [black square in Fig. 1(b)] even though they also see ice $I$ clusters in their simulations [18]. For mW water at 1 bar, however, we [5] get consistent results with those obtained by Li et al. [43] with FFS and by Russo et al. [39] with umbrella sampling, but not fully consistent with those of Haji-Akbari et al. at $\Delta T = 40$ K with FFS [44]. For mW at 2000 and 5000 bar, we also get consistent results with our own brute force calculations of $J$ at high supercooling (see SM [9]). More work is needed to clarify the discrepancies between the different values of $J$ reported for the mW and TIP4P/Ice water models. Another argument that strongly supports the seeding approach is that it successfully predicts nucleation rates and interfacial free energies at coexistence for the well-characterized hard sphere, Lennard Jones and Tosi-Fumi Sodium Chloride models [5].

To conclude, we find that the increase of the ice $I$-water interfacial free energy is the main reason for the decelerating effect of pressure on ice nucleation (at least up to 2000 bar). Our work provides a physical explanation to the high pressure freezing techniques used in the preservation of food and biological samples [2,3] and may serve as guidance to experimentally obtain amorphous water [10,11,45] or to probe metastable supercooled water in no man's land [11,26,46]. Moreover, our work strongly supports that ice nucleates via ice $I$ clusters [40–42], which is crucial information to interpret ice nucleation experiments [28,33].

This work was funded by Grant No. FIS2013/43209-P of the Ministerio de Economia y Competitividad (MEC), the Marie Curie Career Integration Grant No. 322326-COSAAC, and the Universidad Complutense de Madrid (UCM)/Santander Grant No. 910570. C. V. and E. S. acknowledge financial support from a Ramon y Cajal fellowship. J. R. E. acknowledges financial support from the Formacion de Personal Investigador (FPI) Grant No. BES-2014-067625. Calculations were carried out in the supercomputer facility Tirant from the Spanish Supercomputing Network (Red Española de Simulación) (Grant No. QCM-2015-1-0028).